\documentclass[twocolumn,superscriptaddress,floatfix,prl,showpacs]{revtex4-1}
\usepackage{graphicx}
\usepackage{amsmath}
\usepackage{color}
\usepackage{braket}

\begin{document}
\title{Electron-phonon coupling in undoped cuprate YBa$_2$Cu$_3$O$_6$
estimated from Raman and optical conductivity spectra}

\author{D. Farina}
\affiliation{Scuola Normale Superiore, Piazza dei Cavalieri 7, I-56126, Pisa, Italy}
\affiliation{Istituto Italiano di Tecnologia Center for Nanotechnology Innovation @NEST Piazza San Silvestro 12, 56127 Pisa , Italy}

\author{G. De Filippis}
\affiliation{SPIN-CNR and Dip. di Scienze Fisiche - Universit\`a di Napoli Federico II - I-80126 Napoli, Italy}

\author{A.~S.~Mishchenko}
\affiliation{RIKEN Center for Emergent Matter Science (CEMS),  
2-1 Hirosawa, Wako, Saitama, 351-0198, Japan}
\affiliation{NRC ``Kurchatov Institute", 123182, Moscow, Russia}

\author{N.~Nagaosa}
\affiliation{RIKEN Center for Emergent Matter Science (CEMS),
2-1 Hirosawa, Wako, Saitama, 351-0198, Japan}
\affiliation{Department of Applied Physics, The University of Tokyo,
7-3-1 Hongo, Bunkyo-ku, Tokyo 113, Japan} 

\author{Jhih-An Yang}
\affiliation{Department of Physics, University of Colorado-Boulder Boulder, CO 80309, USA}

\author{D. Reznik}
\affiliation{Department of Physics, University of Colorado-Boulder Boulder, CO 80309, USA}
\affiliation{Center for Experiments on Quantum Materials, University of Colorado-Boulder, Boulder, Colorado, 80309, USA}

\author{Th. Wolf}
\affiliation{Institute for Solid State Physics, Karlsruhe Institute of Technology, D-76021 Karlsruhe, Germany}

\author{V. Cataudella}
\affiliation{SPIN-CNR and Dip. di Scienze Fisiche - Universit\`a di Napoli Federico II - I-80126 Napoli, Italy}


\begin{abstract}
We study experimentally the Raman response of the undoped high-T$_c$ parent compound YBa$_2$Cu$_3$O$_6$, and give a unified theory of the two-magnon Raman peak and optical conductivity based on the Hubbard-Holstein model with electron-phonon coupling (EPC). The Hubbard model without EPC can qualitatively account for the experimentally observed resonance of the Raman response, but only the Hubbard-Holstein model (i) reproduces asymmetry of the Raman spectrum, (ii) validates experimental visibility of the two-magnon peak, and (iii) predicts the correct shape and energy of the lower edge of the charge transfer gap in optical conductivity. Comparison of experiments with the theory gives the EPC strength $\lambda = 0.6$. This result convincingly indicates the vital role of EPC in high-Tc cuprates providing a clue to the mechanism of high-T$_c$.
\end{abstract}
\maketitle

High critical temperature (high-$T_c$) superconductivity is the phenomenon whose understanding 
is not only a challenge for descriptive power of the modern theoretical concepts but also bears 
immense importance for potential numerous applications in many fields of innovative technology.  
In spite of enormous efforts to understand the physics of high-$T_c$ there is no adopted opinion 
on the driving forces leading to the superconducting transition yet \cite{LeeNagaosa2006_RevMod}.
Moreover, there is even no consensus on which types of interactions are crucial for the description 
of the normal state of high-$T_c$ compounds. 
It has been adopted by most that the unusual superconductivity of high-$T_c$ 
compounds cannot be described by conventional Bardeen-Cooper-Schrieffer (BCS) mechanism
based on the electron-phonon coupling (EPC) and, hence, one has to assume an important role of the 
electron-electron interaction (EEI).
The emphasis on the EEI in a majority of considered theoretical concepts puts the EPC out of 
the picture leaving an impression that the EPC does not play any role in the physics of  
high-$T_c$ materials. 

However, it has been shown by recent studies that the EPC manifests itself in many phenomena
\cite{Dever95,Khal97,RoGu_EPI_tJ,tJ_ph_2004,RoGu_EPI,RoGu_ARPES,Sangio,Tdep_tJ_2007,OC_tJ_2008,Fausti_NatComm_2014} 
and it was concluded that one needs both EEI and EPC to describe high-$T_c$ materials \cite{RoGu_CM,UFN_2009}.
The main class of unconventional superconductors are cuprates whose parent undoped compounds 
are in the Mott insulating antiferromagnetic (AF) state. 
Doping of these compounds by holes destroys AF state and induces superconductivity.
Recent theoretical studies based on nonperturbative approaches established that the 
EPC is strongly reflected in spectroscopy of undoped and weakly doped compounds though its manifestations weaken with hole doping \cite{OC_tJ_2008,tJ_ARPES_2011}.   

Hence, to address the role of EPC, we focus on 
undoped compounds where EPC is manifested
most clearly, as the basis to construct the theoretical 
model describing cuprates. This enables the quantitative 
estimate of the strength of EPC.

To verify the importance of EPC, we calculated the polarization-resolved two-magnon Raman spectrum 
(RS) and optical conductivity (OC) of the undoped ($\delta=0$) 
YBa$_2$Cu$_3$O$_{6+\delta}$ (YBCO) which is one of the reference high-$T_c$ materials. 
Our calculations show that solely EEI-based description, using model parameters required to 
describe angle resolved photoemisson spectra of high-$T_c$ compounds, is not
successful whereas inclusion of rather substantial EPC not only improves description of both 
RS and OC but provides a unique possibility to describe both experimental 
responses within the same unified model.    

{\it Model.} EEI is introduced in the framework of the extended two-dimensional effective 
one-band  
Hubbard model which has been derived elsewhere \cite{feiner1996effective,simon1997optical} from the more general three band  description. In addition, we will take into account the coupling between the charge carriers and the vibrational modes of the lattice. The Hamiltonian is:   
\begin{equation}
H=H_{\mbox{\scriptsize H}}+H_{\mbox{\scriptsize PH}}+H_{\mbox{\scriptsize EPC}} \; .
\label{h_tot}
\end{equation}
The first term describes pure electronic system with the strong on site Hubbard Coulomb 
repulsion $U$, nearest neighbor coupling constant $V$, and next nearest neighbor constant $V'$ 
\begin{figure}[thb]
\includegraphics[scale=0.4,width=0.99\columnwidth]{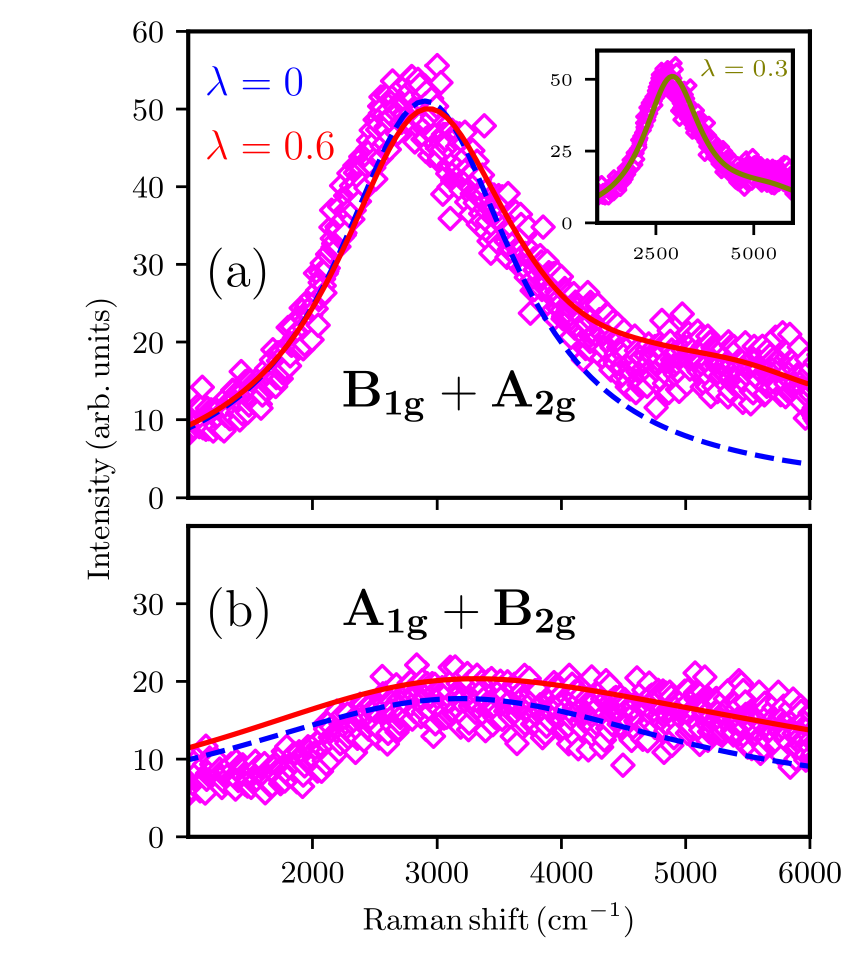}
\caption{\label{fig:1} (color online) 
Calculated Raman signal in (a) B$_{1g}+$A$_{2g}$ and (b) A$_{1g}+$B$_{2g}$ symmetries
without (dash blue line) and with (solid red line) EPC
($\lambda=0.6$). In the inset calculated Raman signal in B$_{1g}+$A$_{2g}$ symmetry
at $\lambda=0.3$.
Experimental data shown by diamonds were obtained on a single crystal sample of insulating 
YBa$_{2}$Cu$_{3}$O$_{6+x}$ using 3.05eV incident laser energy on the McPherson 
triple Raman spectrometer at 300K.
}
\end{figure}
\begin{equation}
\begin{aligned}
H_{\mbox{\scriptsize H}}=-t \sum_{i, \delta, \sigma} c_{i+\delta, \sigma}^{\dagger} c_{i, \sigma} + U \sum_i n_{i, \uparrow} n_{i, \downarrow} \\
+ V \sum_{i \delta \sigma \sigma'} n_{i+\delta \sigma} n_{i \sigma'} + V' \sum_{i \delta' \sigma \sigma'} n_{i+\delta' \sigma} n_{i \sigma'}.
\end{aligned}\end{equation}
\begin{figure}[tbh]
\includegraphics[scale=0.4,width=0.97\columnwidth]{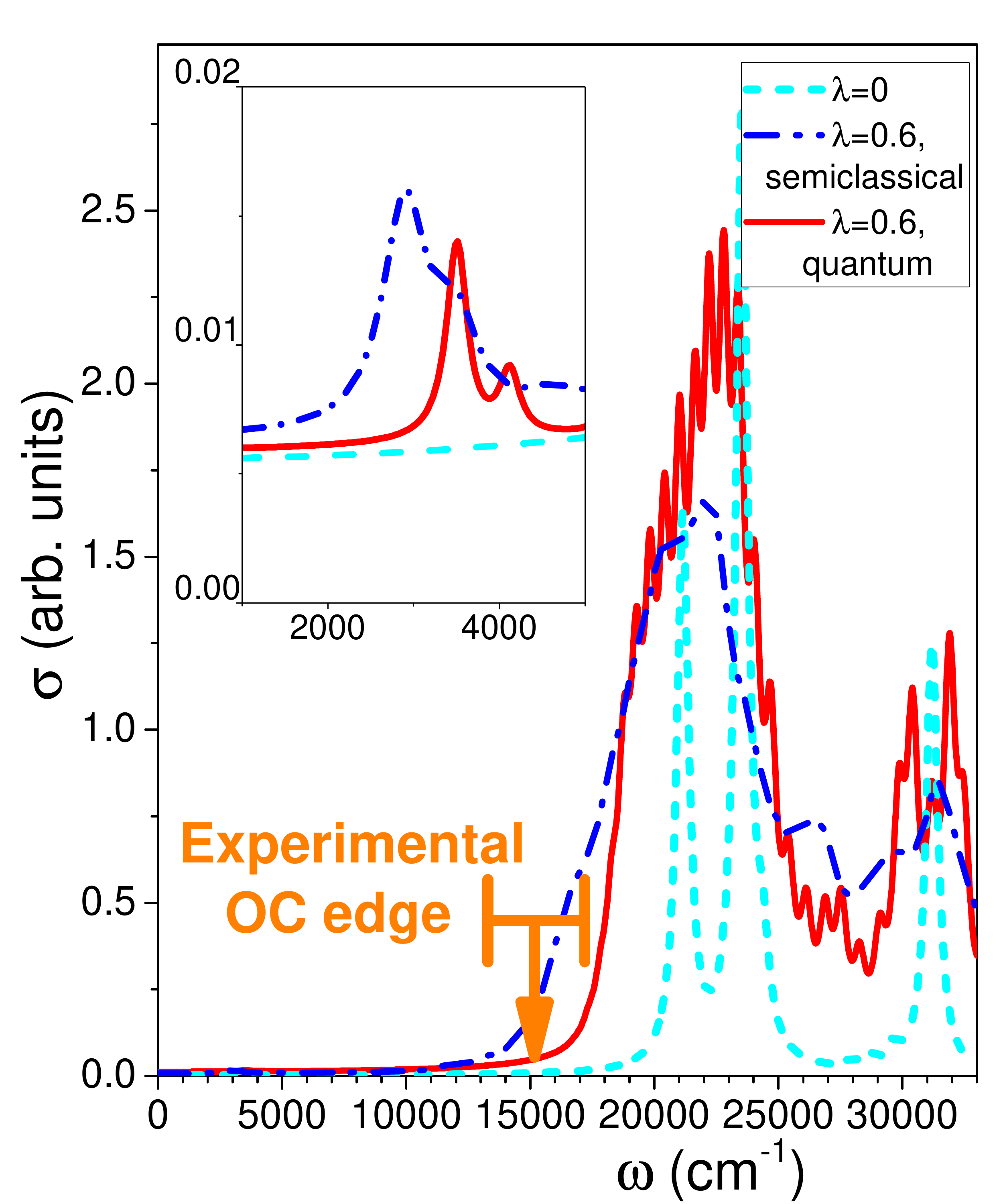}
\caption{\label{fig:2} (color online) 
Theoretical optical conductivity at $\lambda=0$ (cyan dashed line) and 
at $\lambda=0.6$ within the semi-classical (blue dash-dotted line)  
and quantum (red solid line) approaches for phonons. 
The range of OC edge observed in experiments 
(12100cm$^{-1}$-16100cm$^{-1}$) is given by orange error-bar with
orange arrow.
Inset shows comparison of  OCs in the low-energy window.
}
\end{figure}

The vibrational subsystem is described by the out-of-plane dispersionless phonon of 
apical oxygen ions 
in YBCO: 
\begin{equation}
H_{\mbox{\scriptsize PH}}= \omega_0 \sum_i a_i^{\dagger} a_i \; .
\end{equation}
These couple to charge fluctuations  
\begin{equation}
H_{\mbox{\scriptsize EPC}}=g \omega_0  \sum_{i} \left( n_i - 1 \right) (a_i^{\dagger} + a_i) 
\end{equation}
by on-site Holstein type EPC whose strength is characterized by dimensionless 2D coupling
constant $\lambda=g^2 \omega_0/(4 t)$.
The values of the parameters entering Eq.(\ref{h_tot}) have been chosen in agreement 
with the literature \cite{tohyama2002resonant,seibold1998striped,raghu2012effects}. In the
present paper we adopt
$t=0.36 eV$, $U=10 t$, $\omega_0 = 0.2 t$,  $V=0.2U$,  $V'=0.1U$
($V$ and $V'$ have been chosen by assuming a Youkawa-like
electron-electron potential)\cite{Supplement}.
The antiferromagnetism is controlled by the
Heisenberg exchange energy, $J=4 t^2/U$, that turns out to be $J=0.4 t$.
To calculate the optical response we used exact diagonalization of
small systems with semi-classical
phonon in adiabatic approximation (Raman and OC) and with quantum phonons (OC)
(see the Supplemental Material \cite{Supplement}).
To learn about the importance of EPC we compared theoretical description with
($\lambda=0.6$) and without ($\lambda=0$) EPC. We emphasize that
the value $\lambda=0.6$ restores the correct behavior
of OC at very low dopings \cite{Hub_Time_Depend_2012}. 

\begin{figure}[bth]
\includegraphics[scale=0.4,width=0.9\columnwidth]{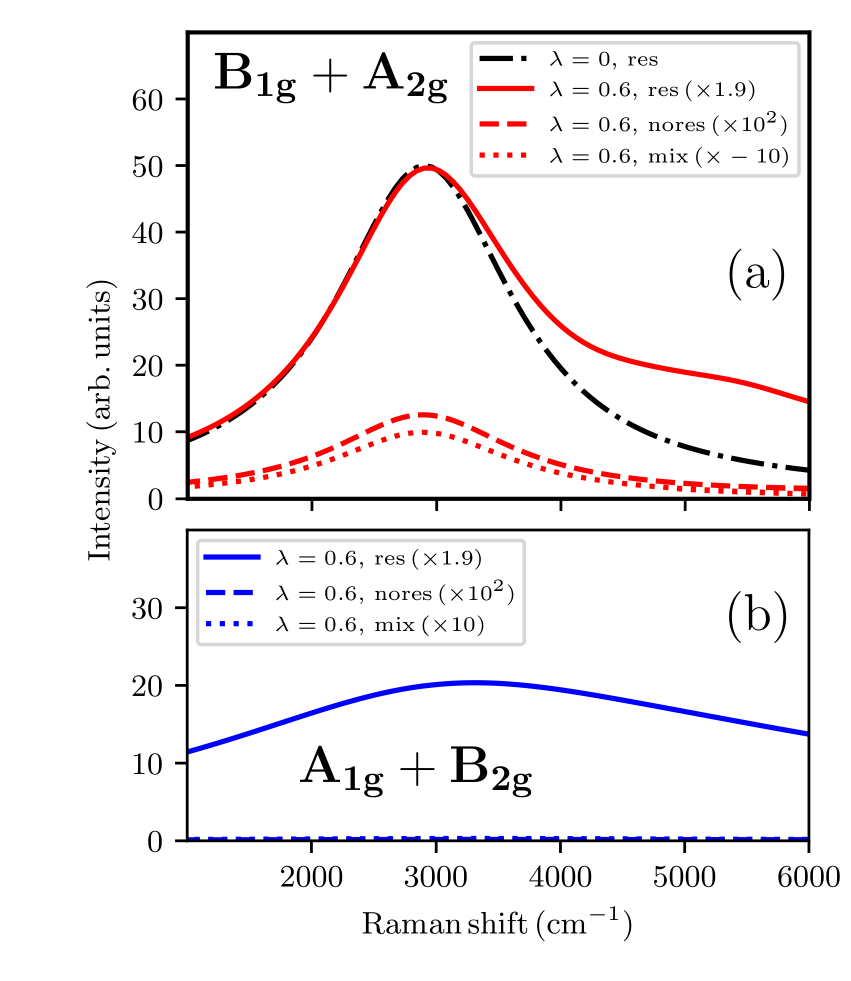}
\caption{\label{fig:3} (color online) 
Resonant (solid line), mixed (dotted line), and non-resonant (dash line)  contributions 
to the Raman response in (a) $B_{1g}+A_{2g}$ and (b) $A_{1g}+B_{2g}$ symmetries 
with electron-phonon interaction  ($\lambda=0.6$) for incoming laser frequency 
$\omega_L=25250\mbox{cm}^{-1}$. 
Resonant contribution at $\lambda=0$ is given in panel (a) by black dotted line.}
\end{figure}

\begin{figure}[bth]
\includegraphics[scale=0.4,width=0.9\columnwidth]{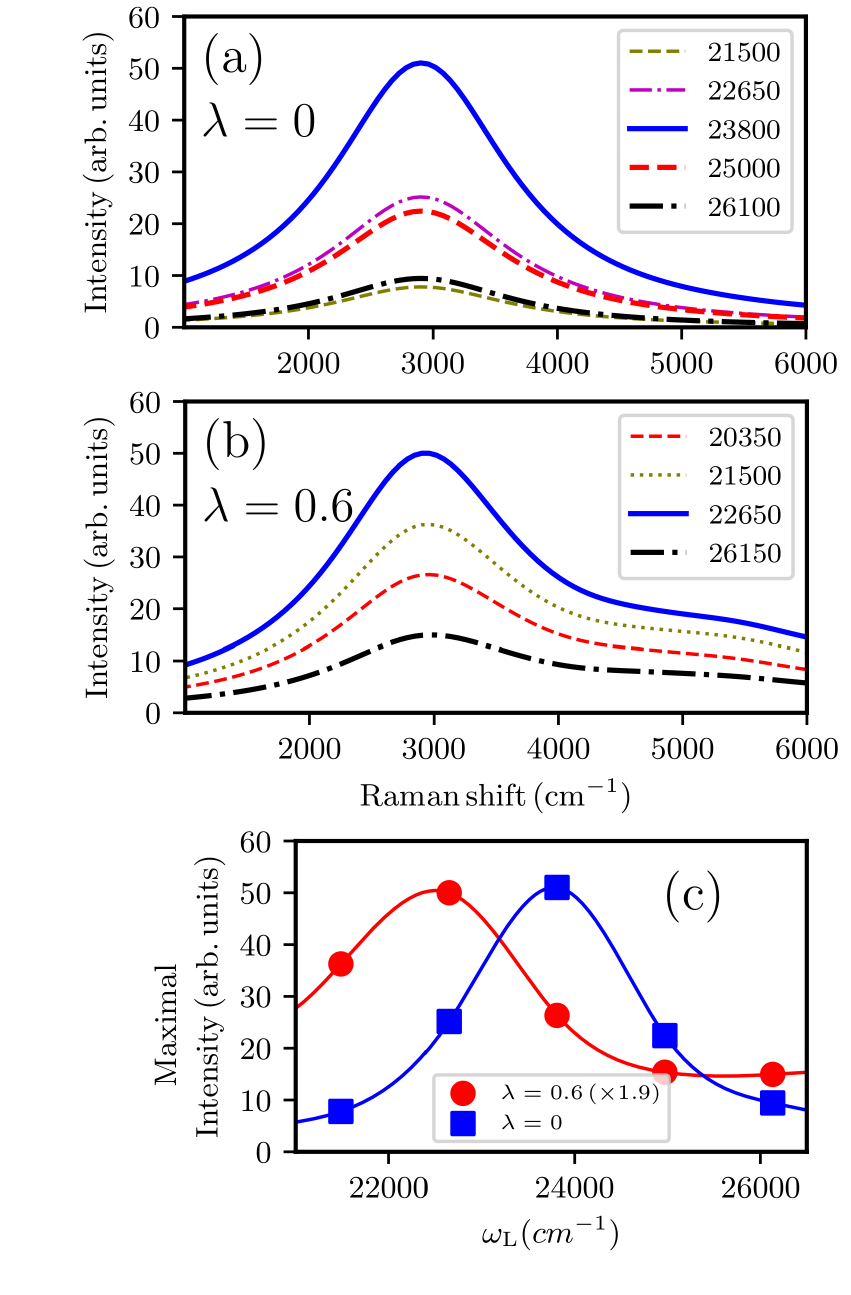}
\caption{\label{fig:4} (color online) 
Resonant behavior of Raman response in $B_{1g}+A_{2g}$ symmetry 
(a) without and (b) with electron-phonon coupling ($\lambda=0.6$). 
The incoming laser frequencies $\omega_L$ in $\mbox{cm}^{-1}$
are given in figure legends. 
 Panel (c) shows dependence of the maximal Raman signal  intensity
 on the incoming laser frequency $\omega_L$.}
\end{figure}

 {\it Properties where EPC is crucial for describing experiments.}
 In polarization-resolved Raman response we focused on the bimagnon peak (2M-peak) which is 
 located at Raman shift $\omega = \omega_L - \omega_S$ around $3000cm^{-1}$
\cite{Reznik_1992_effect, Yoshida_1992_TwoMag,Blumberg_1996_ResTwoMag}. 
This shift is the energy loss between incoming laser light with frequency $\omega_L$ and 
outgoing light frequency  $\omega_S$.
 The adopted explanation of the nature of 2M-peak is given by Chubukov-Frenkel theory
 \cite{Chubukov_1995_resonant}. 
 According to this theory the incident light $\omega_L$ creates electron-hole pair through 
 the electronic gap which is followed by emission of two bound magnons with the  
 opposite momenta decreasing electron-hole pair energy by $\omega$. 
 Consequent recombination leads to emission of light with smaller frequency $\omega_S$ 
 than incoming light by Raman shift $\omega$.    
This process is resonant and the intensity of 2M-peak increases when either $\omega_L$ matches 
the upper edge or when $\omega_S$ matches the lower edge of the electronic gap 
\cite{morr,hanamura}.
 
 The energy of 2M-peak is mainly determined by EEI whereas, as shown in Fig.~\ref{fig:1}, 
 introduction of EPC significantly improves similarity of the theoretical description of the 
Raman response to that measured in experiment.
 The bimagnon excitation is most pronounced in the Raman response in the $B_{1g}+A_{2g}$ 
channel where $B_{ng}$ and $A_{ng}$ are irreducible representations of the YBCO crystal 
point group $D_{4h}$. 
This is experimentally detectable by Raman spectroscopy in the $x^\prime y^\prime$ 
polarization configuration when the incoming ${\bf e}_L$ and outgoing ${\bf e}_S$ 
photon polarizations are perpendicular to each other and oriented at $45 ^\circ$ with 
respect to the 2D lattice bonds. 
The complementary symmetry $A_{1g}+B_{2g}$ or $x^\prime x^\prime$ is 
experimentally obtained by rotating $\mathbf{e}_S$ to make it parallel to 
$\textbf{e}_L$ along the $x^\prime$ direction. 

Without the EPC, in severe contrast with experimental data 
\cite{Blumberg_1996_ResTwoMag,CoMa}, the
theoretical 2M-peak in the $B_{1g}+A_{2g}$ channel is perfectly symmetric with respect
to the bimagnon energy $\omega_{2M} \sim 2.5J \approx 3000$cm$^{-1}$.
Inclusion of the EPC cures this discrepancy between the theory and experiment and
quantitatively
reproduces the asymmetry, see Fig.~\ref{fig:1}a (inset in Fig.~\ref{fig:1}a points out
that the best agreement with experimental observations is obtained at $\lambda=0.6$).
On the other hand
EPC plays a minor role in the $A_{1g}+B_{2g}$ symmetry (see Fig.~\ref{fig:1}b). However,
also in this case,
the agreement with experimental observations is improved by the inclusion
of charge lattice coupling.

As for the OC, the very visibility of the 2M-peak in the theoretical description is the consequence of
EPC, see inset in Fig.~\ref{fig:2}.  
Intensity of 2M-peak in rigid YBCO is suppressed by the point inversion symmetry 
of the unit cell and only EPC makes the 2M-peak visible in principle because phonons 
break this high symmetry.
We note that although the intensity of the bimagnon peak in OC is orders of magnitude 
weaker than the spectral weight above the gap 
\cite{Lorenzana_1995_resonant,Hub_Time_Depend_2012}, the very presence of this signal,
observed in experiment \cite{perkins1993mid},  
is an unambiguous proof of the importance of EPC.

Inclusion of EPC provides other important improvements concerning
the shape and the value of the gap in the OC, see Fig.~\ref{fig:2}.
Indeed, in the ground state at half-filling,
Coulomb repulsion forces the electrons to localize,
freezing charge fluctuations. In such a state,
the coupling between localized fermions and bosonic excitations,
which is mediated by charge dynamics, is strongly suppressed.
On the other hand, at the edge of OC,
where holons and doublons are formed, charge
lattice coupling becomes relevant.
EPC moves the edge of the OC to lower energies building up
a characteristic tail just as reproduced by experiments
\cite{PhysRevB.42.6342,PhysRevB.48.7545,cooper}. Here the main role is played by
excitons strongly dressed by phonons.
On the other hand, the effect of phonons is
less important well inside the absorption band where no polaron
can be formed.
Nevertheless a simple broadening of the peaks is observed above the charge transfer
gap too.

{\it Properties where EPC is not crucial but plays a big role.}
Here we discuss the properties, which are not substantially modified by the inclusion 
of EPC. 
To this aim one has to consider the structure of Raman response and OC. 
The exact eigenstates representation for the polarization-resolved electronic Raman spectrum
at zero temperature as a function of the Raman shift $\omega \equiv \omega_L - \omega_S > 0$, is given by \cite{Shastry_1990_theory,Devereaux_2007_inelastic}
\begin{equation}
\begin{aligned}
I_{Raman}(\omega; \textbf{e}_L, \textbf{e}_S)\propto 
\; \; \;  \; \; \; \; \; \;  \; \; \;  \; \; \; \; \; \; \; \; \; \; \\
\frac{\omega_S}{\omega_L} \sum_f |\langle{\Psi_f} |  \mathbf{e}_S^\dagger M \mathbf{e}_L
| {\Psi_0} \rangle  |^2 \times \Im{\frac{1}{\omega-E_f+E_{0}-i \epsilon} } 
\end{aligned}
\end{equation} 
where ${\bf e}_L$ and ${\bf e}_S$ are polarizations of the incoming and outgoing light, 
$E_0$ and $| {\Psi_0} \rangle$ are energy and wave function of the ground state, and 
$E_f$ and $| {\Psi_f} \rangle$ are energies and wave functions of the final states.
The matrix elements of the Raman scattering tensor operator
\begin{equation}
\begin{aligned}
\langle{\Psi_f} | M_{lm} | {\Psi_0} \rangle = \langle{\Psi_f} | \tau_{lm} | {\Psi_0} \rangle +
\; \; \;  \; \; \; \; \; \;  \; \; \; 
\\
\sum_{r}
\left\{
\frac{\langle{\Psi_f}| j_l | {\Psi_r} \rangle \langle{\Psi_r}| j_m |{\Psi_0}\rangle}
{\omega_L + E_{0} -E_r -i\eta}
-
\frac{\langle{\Psi_f}| j_m|{\Psi_r} \rangle \langle{\Psi_r} |j_l|{\Psi_0} \rangle}
{\omega_L + E_{r} -E_f -i\eta}
\right\}
\end{aligned} 
\label{eq:raman_cross_sec}\end{equation}
contain two contributions.
The first term is non-resonant, is determined by the Raman stress tensor operator
$\tau_{lm}$ (see the Supplemental Material \cite{Supplement}), and is
insensitive to the incoming photon frequency $\omega_L$.
The second one strongly depends on the frequency $\omega_L$, which can resonate only
with the difference of the energies of the intermediate $|{\Psi_r} \rangle$ and
ground states, $E_r-E_0$, because of total energy conservation. The
resonant term contains the components of the current operator $j_l$
(see the Supplemental Material \cite{Supplement}),
and the intermediate states $|{\Psi_r} \rangle$. 
The particular structure of the Raman response makes it much more sensitive to 
symmetry breaking than OC \cite{Gallais_2016_charge,Massat_2016_charge}, 
namely to degenerate eigenvalues of the system Hamiltonian.  
The expression for OC reads as follows
\begin{equation}
\begin{aligned}
\Re{ \sigma^{reg}_{xx}(\omega)}=\sum_{n\neq 0} \frac{|\langle{\Psi_n}| j_x |{\Psi_0}\rangle|^2 }{E_n-E_0} 
\times 
\; \; \;  \; \; \; \; \; \;  \; \; \;
\\
\Re{\left[ i \left( \frac{1}{\omega + i \delta -E_n+E_0} - \frac{1}{\omega + i \delta + E_n-E_0} \right)\right]} \; .
\end{aligned}
\end{equation}

The dominant contribution to the Raman scattering, both with and without EPC, comes from the 
resonant contribution, see Fig.~\ref{fig:3} where different terms of the Raman response are
compared for $\lambda=0.6$. 
Fig.~\ref{fig:3}a compares resonant contributions at $\lambda=0.6$ and
$\lambda=0$. 
One can conclude that  the EPC is responsible for the experimentally observed asymmetry of 
the 2M-peak in $B_{1g}+A_{2g}$ symmetry and that EPC mostly affects the resonant 
contribution. This stems from the observation that the resonant
contribution involve states at the edge of OC, where
holons and doublons are formed and EPC plays a significant role.  

The resonant behavior is observed both with and without EPC (Fig.~\ref{fig:4}).
In both cases, comparing Fig.~\ref{fig:4}c and Fig.~\ref{fig:2}, one concludes that 
the resonant contribution is maximal when the incident frequency $\omega_L$ matches 
the maximum of the OC above the charge transfer gap. 
However, the resonance is much broader when EPC is included, which is closer to the 
experiment \cite{TwoMag94}.

We emphasize that RS are often discussed in literature within the framework of the 
Fleury-London theory 
based on the coupling between the light and the spin system in the Heisenberg model 
\cite{sandvik1998numerical}. However these approaches do not reproduce resonant scattering 
occurring when the frequency of the incoming light is comparable to the charge transfer gap. 
In order to recover the experimental resonance one has to take into account the full Hubbard model \cite{tohyama2002resonant}.    

{\it Conclusion.} 
We compared the capabilities of the extended Hubbard and extended Hubbard-Holstein models to give a
unified description of the Raman response and optical conductivity 
of high-$T_c$ superconductors on the example of the prototypical undoped compound ($\delta=0$) 
YBa$_2$Cu$_3$O$_{6+\delta}$.
We showed that both models can explain the experimentally observed resonant
nature of the Raman response.  
However,  we found that the extended Hubbard-Holstein model, including electron-phonon coupling,
gives a better description of experimental data. 
First, the Hubbard-Holstein model, in contrast with the pure Hubbard model, reproduces experimentally observed 
asymmetry of the Raman spectrum.
Second, the presence of the electron-phonon coupling is manifested in experimental visibility of the two-magnon 
peak in optical conductivity.
Finally, the Hubbard-Holstein model predicts correct positions of peaks both in the 
Raman response and optical conductivity with the same parameters, i.e. it provides a unified description of two spectral
properties in a situation where pure Hubbard model fails.    

This work was funded by ImPACT Program of Council for Science, Technology and
 Innovation (Cabinet Office, Government of Japan). Work at the University of Colorado
was supported by the NSF under Grant No. DMR-1709946.




%
%

\end{document}